\documentclass{article}
\usepackage[pdftex]{graphicx}

\usepackage[preprint]{neurips_2025}

\usepackage[utf8]{inputenc} 
\usepackage[T1]{fontenc}    
\usepackage{url}            
\usepackage{booktabs}       
\usepackage{amsfonts}       
\usepackage{nicefrac}       
\usepackage{xcolor}         

\usepackage{makecell}
\usepackage{amsmath}

\title{QFGN: A Quantum Approach to High-Fidelity Implicit Neural Representations}

\author{
  Hongni Jin, Gurinder Singh, and Kenneth M. Merz, Jr. \\
  Center for Computational Life Sciences, Lerner Research Institute \\
  Cleveland Clinic, Cleveland, OH 44106\\
  \texttt{\{jinh2,singh12,merzk\}@ccf.org} \\
}

\begin{document}

\maketitle

\begin{abstract}
    Implicit neural representations have shown potential in various applications. 
    However, accurately reconstructing the image or providing clear details via image super-resolution remains challenging. This paper introduces Quantum Fourier Gaussian Network (QFGN), a quantum-based machine learning model for better signal representations. 
    The frequency spectrum is well balanced by penalizing the low-frequency components, leading to the improved expressivity of quantum circuits. 
    The results demonstrate that with minimal parameters, QFGN outperforms the current state-of-the-art (SOTA) models. 
    Despite noise on hardware, the model achieves accuracy comparable to that of SIREN, highlighting the potential applications of quantum machine learning in this field. 
\end{abstract}

\section{Introduction}

Implicit Neural Representations (INRs) have emerged as a powerful tool, offering a paradigm shift in how images and signals are modeled.\citep{sitzmann2020implicitneuralrepresentationsperiodic,tancik2020fourierfeaturesletnetworks,pmlr-v250-sideri-lampretsa24a}
Instead of storing data as discrete arrays, INRs represent complex structures as continuous functions parameterized by neural networks, mapping spatial coordinates directly to signal values such as intensity or semantic labels.
This approach enables resolution-independent modeling, compact storage, and adaptive sampling—all of which are especially beneficial for handling large-scale volumetric scans. 
For example, INRs are beginning to reshape medical imaging by offering novel solutions for key challenges such as high-resolution image reconstruction, precise anatomical alignment through registration, and memory-efficient 3D modeling. 
Early applications have demonstrated that INRs can capture fine structural detail, adapt flexibly to varying image resolutions, and reduce computational overhead—making them a promising foundation for next-generation imaging pipelines.\citep{molaei2023implicitneuralrepresentationmedical,wang2025neuralradiancefieldsmedical,wei2024imedsamimplicitmedicalimage}

While INRs have demonstrated strong empirical performance, a deeper understanding of their internal mechanisms remains essential for robust and principled application.
Benbarka et al.\citep{benbarka2021seeingimplicitneuralrepresentations} proposed a novel interpretation of INRs as Fourier series, offering a theoretical bridge between coordinate encodings and signal representation.
They show that a perceptron with a Fourier-mapped input is mathematically equivalent to a truncated Fourier series, where the network’s weights serve as the series coefficients. 
This perspective hints why periodic activation functions (like sine in SIRENs\citep{sitzmann2020implicitneuralrepresentationsperiodic}) or sinusoidal encodings\citep{tancik2020fourierfeaturesletnetworks} help multilayer perceptrons (MLPs) capture high-frequency components: they effectively mimic classical signal decomposition.
However, this Fourier-like behavior must be manually imposed in classical neural networks through architectural choices and input encodings.

Quantum machine learning (QML) is an emerging field at the intersection of quantum computing and AI, aiming to harness the computational advantages of quantum systems for data-driven tasks.
The QML models are typically implemented using parameterized quantum circuits (PQCs), where quantum gates are controlled by trainable parameters, and the output is obtained by measuring quantum observables.
These circuits operate in a high-dimensional Hilbert space, allowing them to represent complex functions with relatively few parameters.\citep{singh2025benchmarkingmedmnistdatasetreal}
Beyond potential quantum speedups, recent work\citep{PhysRevA.103.032430} has shown that PQCs exhibit an inherent Fourier structure, indicating that the output of a quantum circuit is mathematically a Fourier series in the parameters.
This intrinsic property aligns naturally with the goals of INRs, suggesting that QML may serve not only as a new computational strategy but also as a native architecture for function approximation in INR tasks.

In this work, we propose Quantum Fourier Gaussian Network (QFGN), a novel hybrid classical-quantum model for INRs.
The classical layers are mainly designed to provide a uniformly distributed frequency spectrum including low-frequency and high-frequency components to mitigate the spectral bias\citep{rahaman2019spectralbiasneuralnetworks}.
These Fourier features are used as data encoding in quantum circuits to fit implicit functions. The results validate that QFGN shows remarkable performance on medical image reconstruction and super-resolution. 
Moreover, even with noise on quantum hardware, QFGN still outperforms some SOTA models.
In summary, our contributions are (1) we introduce QFGN, a quantum-involved model for continuous function fitting and validate the quantum advantage of this model; (2) we design a novel classical layer to penalize the amplitude of low-frequency components; (3) we provide a comprehensive analysis of the effect of error mitigation and hardware-specific variability on QML models implemented on hardware and validate the potential of QML in the Noisy Intermediate-Scale Quantum(NISQ) era. 

\section{Related Work}
\subsection{Implicit Neural Representations}
INRs offer an alternative representation of continuous signals by modeling them as implicit continuous functions parameterized by neural networks, rather than relying on discrete grid-based representations.
Recent work has demonstrated the remarkable performance and memory-efficient properties of INRs across a wide range of tasks. 
For example, INRs have been successfully applied in 2D/3D image representation\citep{saragadam2023wirewaveletimplicitneural,xie2022dinerdisorderinvariantimplicitneural}, super-resolution\citep{wu2021iremhighresolutionmagneticresonance,Wu_2023}. 
They are also effective in occupancy volume representation\citep{peng2020convolutionaloccupancynetworks}, allowing continuous and compact modeling of geometry. In view synthesis\citep{mildenhall2020nerfrepresentingscenesneural,zhang2024neuralimplicitrepresentationhighly}, INRs such as Neural Radiance Fields (NeRF) enable photorealistic rendering from sparse viewpoints. 
Additionally, INRs are increasingly used in physics-informed modeling\citep{Pezzoli_2024}, audio signal modeling\citep{szatkowski2024hypersoundgeneratingimplicitneural} and video compression\citep{kwan2023hinerv}, where continuous representations are essential for smooth, real-time interactions. 

INRs aim at learning a continuous mapping from coordinates to signal values, \( f: \mathbb{R}^{d_{\text{in}}} \to \mathbb{R}^{d_{\text{out}}} \) 
such that: \( f(\mathbf{x}) \approx \mathbf{y} \), where 
\( \mathbf{x} \in \mathbb{R}^{d_{\text{in}}} \) is a coordinate, 
\( \mathbf{y} \in \mathbb{R}^{d_{\text{out}}} \) is the target signal value and 
\( f \) is a MLP, \( f(\mathbf{x}) = \sigma(W\mathbf{x} + \mathbf{b}) \), 
where \( \sigma \) is a nonlinear activation function, 
\( W \) and \( \mathbf{b} \) are trainable weights.
In the early development of INRs, ReLU was predominantly employed as the activation function.\citep{shenouda2024relussufficientlearningimplicit} 
However, due to the inherent spectral bias\citep{rahaman2019spectralbiasneuralnetworks}—-where lower-frequency components are learned preferentially—ReLU-based models often exhibit significant reconstruction errors when attempting to represent fine-grained or high-frequency signal details. 
To mitigate this bias, several novel activation functions have been proposed, like SIREN\citep{sitzmann2020implicitneuralrepresentationsperiodic}, Wire\citep{saragadam2023wirewaveletimplicitneural} and Finer\citep{liu2023finerflexiblespectralbiastuning}. 
In addition, Fourier feature mapping is another effective approach to enable the learning of broader frequency content in continuous functions. 
It first maps the input coordinates to a higher-dimensional space using predefined Fourier features. 
For example, random Fourier features (RFF)\citep{tancik2020fourierfeaturesletnetworks},
\(
\boldsymbol{\gamma}(\mathbf{x}) = 
\left[
\alpha \cos(2\pi B \mathbf{x}),\ 
\alpha \sin(2\pi B \mathbf{x})
\right]^T
\), uses a random matrix \( B \in \mathbb{R}^{m \times d_{\text{in}}} \), 
where \( \mathbf{m} \) is the number of frequencies, sampled from a Gaussian distribution as the Fourier basis frequencies, and these computed Fourier features are then fed into a MLP to approximate the target function. 

While SIREN and Fourier feature mapping appear structurally different, 
they are in fact deeply connected. Benbarka et al.\citep{benbarka2021seeingimplicitneuralrepresentations} 
demonstrate that a MLP with Fourier-mapped inputs is structurally analogous to a single-hidden-layer SIREN network 
by showing that both architectures ultimately represent functions \( f(\mathbf{x}) \) as weighted sums of sinusoidal components, 
\begin{equation}
f(\mathbf{x}) = \sum_{\mathbf{n} \in \mathbb{Z}^{d_{\text{in}}}} a_{\mathbf{n}} \cos(2\pi \mathbf{n} \cdot \mathbf{x}) 
+ b_{\mathbf{n}} \sin(2\pi \mathbf{n} \cdot \mathbf{x})
\label{eq:fourier-series}
\end{equation}
This structural equivalence highlights that both approaches leverage similar inductive biases rooted in Fourier series representations, providing theoretical justification for the effectiveness of Fourier-based methods in INRs. 

\subsection{Quantum circuits as Fourier series}
\label{sec:qc as Fourier series}
The quantum circuit in a typical QML framework includes two building blocks which are composed of quantum gates.
The data encoding block is designed to encode input data, which usually transforms classical data into quantum states while the trainable block has quantum gates with trainable parameters. 
At the end of a quantum circuit, the qubits are measured multiple times to estimate the expectation value of predefined observables, and the results are used as the QML prediction. 
A parametrized quantum model \( f(\mathbf{x}, \boldsymbol{\theta}) \) with regard to some observables \( O \) is defined as
\begin{equation}
f(\mathbf{x}, \boldsymbol{\theta}) = \langle 0 | U^{\dagger}(\mathbf{x}, \boldsymbol{\theta}) \mathcal{O} U(\mathbf{x}, \boldsymbol{\theta}) | 0 \rangle
\label{eq:quantum-circuit}
\end{equation}
where \( U(\mathbf{x}, \boldsymbol{\theta}) \) is a quantum circuit including both input data \( \mathbf{x} \in \mathbb{R}^{d_{\text{in}}} \) and trainable parameters \( \boldsymbol{\theta} \).
Similar to the framework of multiple layers with multiple neurons in a classical neural network, where data is processed multiple times through multiple MLPs, Perez-Salinas et al.\citep{P_rez_Salinas_2020} proposed a data re-uploading scheme to improve the expressivity of quantum circuits. 
Classical input data is reintroduced multiple times into the quantum circuit at different stages. 
Each re-upload involves encoding the data into the parameters of the quantum gates, enabling the circuit to build more complex architectures, leading to improved efficiency. 
Previous work\citep{PhysRevA.103.032430,zhao2024quantumimplicitneuralrepresentations,mhiri2024constrainedvanishingexpressivityquantum} demonstrates that the mathematical structure of a data re-uploading circuit is indeed a truncated Fourier series, 
\begin{equation}
f(\mathbf{x}) = \sum_{\omega \in \Omega} c_\omega e^{i \omega \cdot \mathbf{x}}
\label{eq:fourier-expansion}
\end{equation}

The data re-uploading circuit takes the general form as
\begin{equation}
U(\mathbf{x}, \boldsymbol{\theta}) = W_\theta^{L+1} S(\mathbf{x}) W_\theta^L S(\mathbf{x}) \cdots S(\mathbf{x}) W_\theta^1
\label{eq:reuploading-circuit}
\end{equation}
which includes \( L \) repetitive layers, and each layer includes one trainable block \( W_\theta \), and one encoding block 
\( S(\mathbf{x}) = e^{-i x H} \), where \( x \) is the input data, and \( H \) is an arbitrary Hamiltonian. We drop 
\( \boldsymbol{\theta} \) in all notations (\( f, U, W \)) for simplicity. Furthermore, this encoding block usually contains 
multiple unitary gates to cover the high-dimensional data, 
\(
S(\mathbf{x}) = e^{-i x_1 H} \otimes \cdots \otimes e^{-i x_{d_{\text{in}}} H}.
\)
We assume that \( H \) is diagonal via eigenvalue decomposition \( H = V^\dagger \Sigma V \), 
where \( V \) is a unitary gate that can be absorbed into the trainable block \( W \), and 
\( \Sigma \) is a diagonal matrix with regard to the eigenvalues of \( H \), including 
\( \lambda_1, \cdots, \lambda_d \). Then the general form of the \( k^\text{th} \) encoding block in the 
re-uploading circuit, \( S^k(\mathbf{x}) \), can be decomposed as
\begin{equation}
[S^k(\mathbf{x})]_{j,j} 
= \exp\left( -i \left( \lambda_{j_1^k}, \cdots, \lambda_{j_{d_{\text{in}}}^k} \right) \cdot \mathbf{x} \right)
= \exp\left( -i \sum_{m=1}^{d_{\text{in}}} \lambda_{j_m^k} x_m \right)
\label{eq:encoding-block-corrected}
\end{equation}
where \( j = (j_1, \cdots, j_{d_{\text{in}}}) \). By using the multi-index notation 
\( j_k = \{j_1^k, \cdots, j_{d_{\text{in}}}^k\} \in \{1, \cdots, d\}^{d_{\text{in}}} \), 
\( S^k(\mathbf{x}) \) can be simplified as 
\(
[S^k(\mathbf{x})]_{j,j} = e^{-i \lambda_{j_k} \cdot \mathbf{x}}.
\)
And the quantum state \( U(\mathbf{x})|0\rangle \) can be represented as
\begin{equation}
[U(\mathbf{x})|0\rangle]_i = \sum_{j_1 \cdots j_L} 
e^{-i(\lambda_{j_1} + \cdots + \lambda_{j_L}) \cdot \mathbf{x}}
\times W^{({L+1})}_{i j_L} \cdots W^{(2)}_{j_2 j_1} W^{(1)}_{j_1}
\label{eq:quantum-state-full}
\end{equation}
The equation is further simplified as
\begin{equation}
[U(\mathbf{x})|0\rangle]_i = \sum_{j_1 \cdots j_L} 
e^{-i \Lambda_j \cdot \mathbf{x}} 
\times W^{({L+1})}_{i j_L} \cdots W^{(2)}_{j_2 j_1} W^{(1)}_{j_1}
\label{eq:quantum-state-simplified}
\end{equation}
where \( \Lambda_j = \lambda_{j_1} + \cdots + \lambda_{j_L} \).
Combining the complex conjugate of the quantum state, the complete quantum model \( f(\mathbf{x}) \) is compactly summarized as
\begin{equation}
f(\mathbf{x}) = \sum_{K, J} a_{K,J} e^{i (\Lambda_K - \Lambda_J) \cdot \mathbf{x}}
\label{eq:fourier-sum}
\end{equation}
\begin{equation}
a_{K,J} = \sum_{i, i'} W^{(1)\dagger}_{1 k_1} W^{(2)\dagger}_{k_1 k_2} \cdots W^{({L+1})\dagger}_{i k_L } 
\ \mathcal{O}_{i i'} \ 
W^{({L+1})}_{i' j_L} \cdots W^{(2)}_{j_2 j_1} W^{(1)}_{j_1 1}
\label{eq:coefficients}
\end{equation}

Eq. (8) indicates that a quantum circuit is indeed a Fourier series, and the frequency spectrum 
of the model is 
\(
\Omega_{KJ} = \{\Lambda_K - \Lambda_J\} = \{\lambda_{k_1} + \cdots + \lambda_{k_L} - (\lambda_{j_1} + \cdots + \lambda_{j_L})\},
\)
which is fully determined by the eigenvalues of the encoding Hamiltonian, while the coefficients \( a_{K,J} \) are controlled by 
the whole circuit, including \( V, V^\dagger \) from the encoding Hamiltonians, the trainable parameters \( \boldsymbol{\theta} \), 
as well as observables \( \mathcal{O} \). The equivalence between Eq. (1) and Eq. (8) derived from Euler’s formula 
indicates that quantum circuits have some structural advantages over neural networks in INRs. Based on this intrinsic 
property of quantum circuits, Zhao et al.\citep{zhao2024quantumimplicitneuralrepresentations} proposed a quantum implicit neural representation network 
(QIREN) to approximate any continuous functions using quantum circuits.

While quantum circuits are structurally Fourier series, the frequency spectrum they have access to 
is finite due to frequency redundancy\citep{landman2022classicallyapproximatingvariationalquantum}. The Pauli rotation gates, such as 
\( R_x, R_y, R_z \), are commonly used as encoding gates, and they share the same eigenvalues, 
\( \lambda = \pm \frac{1}{2} \). For a simple variational circuit with \( L \) encoding gates for 
one-dimensional data, the available \( \Lambda_J \) set is 
\(
\left\{ -\frac{L}{2}, -\frac{L}{2} + 1, \cdots, \frac{L}{2} \right\},
\)
which are all integers if \( L \) is even, or half-integers otherwise. Then only \( 2L + 1 \) unique 
frequencies are available in the frequency spectrum 
\(
\Omega = \{-L, -L+1, \cdots, -1, 0, 1, \cdots, L-1, L\}
\),
however the number of possible values of \( \Lambda_K - \Lambda_J \) is \( 2^{2L} \). This frequency 
redundancy limits the ability of quantum circuits to fit complex functions. To mitigate this issue, several parameter re-scaling strategies have been proposed to expand the frequency 
spectrum by designing encoding gates as 
\(
S(\mathbf{x}) = e^{-i \alpha \mathbf{x} H},
\)
where \( \alpha \) is a scaling parameter to enrich the eigenvalues of the Hamiltonian in each encoding gate\citep{zhao2024quantumimplicitneuralrepresentations,Shin_2023,Peters_2023,Jaderberg_2024,Casas_2023}. 
The introduction of a scaling parameter into the encoding gate can be easily achieved by preprocessing input data: 
\(
\mathbf{x} \rightarrow \vec{\alpha} \mathbf{x},
\)
where \( \vec{\alpha} \) is a multi-dimensional vector to ensure a multi-dimensional Fourier series with 
various frequencies, giving rise to the frequency spectrum
\(
\Omega_{KJ} = \{ \Lambda_K - \Lambda_J \} = 
\left\{ \vec{\alpha}_1 (\lambda_{k_1} - \lambda_{j_1}) + \cdots + 
\vec{\alpha}_L (\lambda_{k_{L-1}} - \lambda_{j_L}) \right\}.
\)
And for a Hamiltonian with \( n \) qubits per input dimension, the quantum circuit can reach 
\( (2nL + 1)^{d_{\text{in}}} \) distinct frequencies with full diversity. This exponential growth of frequency with respect to input dimensionality clearly surpasses the linear growth observed in classical implicit neural representation networks, highlighting the quantum advantage of quantum circuits in INRs.

\section{Quantum Fourier Gaussian Network(QFGN)}
\begin{figure}[htbp]
    \centering
    \includegraphics[width=1\textwidth]{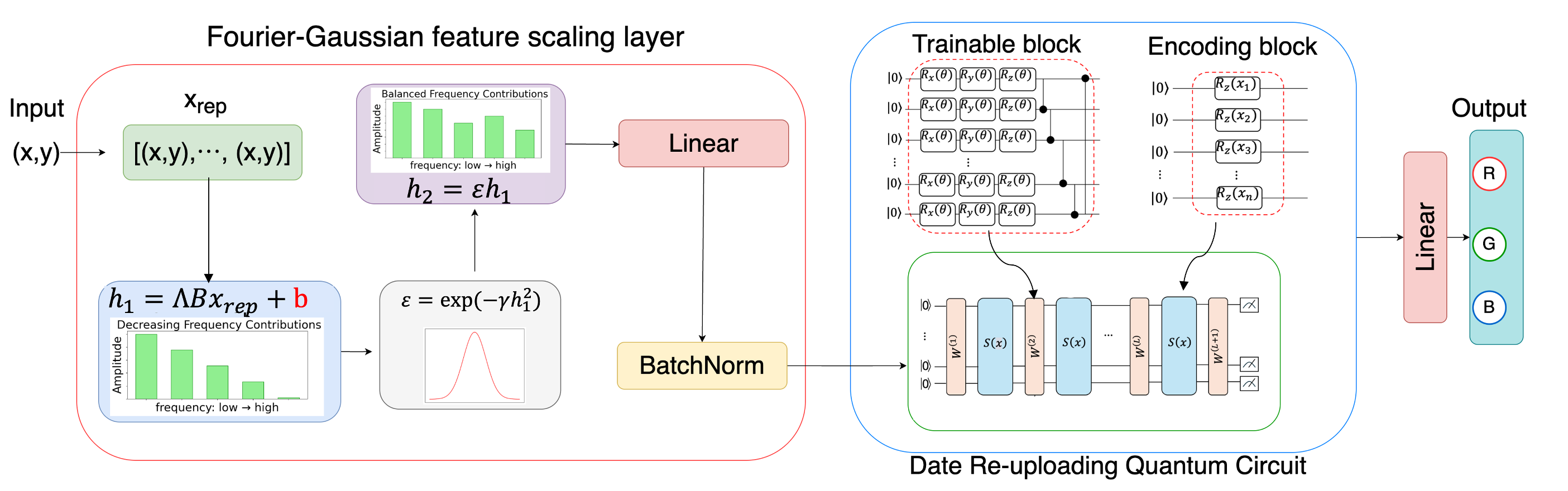}
    \caption{The overall framework of QFGN. In the Fourier part \( h_1 \), only the bias \( b \) is trainable}
    \label{fig:figure1}
\end{figure}

In this section, we introduce Quantum Fourier Gaussian Network(QFGN) to maximize the diversity of the frequency spectrum for quantum circuits.
This classical-quantum hybrid model is designed for INRs, which includes three modules, as shown in Figure~\ref{fig:figure1}. It takes coordinates \( \mathbf{x} \in \mathbb{R}^{d_{\text{in}}} \) as inputs and outputs signals, \( \mathbf{y} \in \mathbb{R}^{d_{\text{out}}} \).

The first module is a Fourier-Gaussian feature scaling layer (FGFS), which expands and scales 
quantum circuit inputs using a structured combination of Fourier basis functions and Gaussian functions. 
Considering the exponential growth of frequencies with respect to the input data dimension \( d_{\text{in}} \), in order to create a multi-dimensional Fourier series, the input coordinate \( \mathbf{x} \in \mathbb{R}^{d_{\text{in}}} \) 
is first repeated \( n \) times with concatenation, giving rise to a high-dimensional input vector 
\( \mathbf{x}_{\text{rep}} \in \mathbb{R}^{nd_{\text{in}}} \). 
\( \mathbf{x}_{\text{rep}} \) is then projected to a set of Fourier bases, derived from the Fourier reparameterization~\citep{shi2024improvedimplicitneuralrepresentation}. 
The basis matrix \( \mathbf{B} \) is composed of elements \( b_{k,j} \), which is given as:
\begin{equation}
b_{k,j} = \cos(w_f s_j + \varphi_p)
\label{eq:basis}
\end{equation}
\begin{equation}
k = (f - 1) \times P + p
\label{eq:k-index}
\end{equation}
where \( w_f \in \{1, 2, \cdots, F\} \) is a frequency array, \( \varphi_p \in \{1, 2, \cdots, P\} \) is a phase array, and the sampling range of \( s \) is chosen as 
\( [-2\pi, 2\pi] \) with the length of \( nd_{\text{in}} \), i.e., the dimension of the concatenated input data. 
The entire matrix of the Fourier basis \( \mathbf{B} \) covers a dense, discrete frequency-phase space, explicitly constructed by iterating first over each frequency 
and then over each associated phase at discrete sampling points:
\begin{equation}
\mathbf{B} = \{ b_{k,j} \mid f = 1, \cdots, F; \ p = 1, \cdots, P; \ j = 1, \cdots, nd_{\text{in}} \} 
\in \mathbb{R}^{(F \times P) \times nd_{\text{in}}}
\label{eq:basis-matrix}
\end{equation}

\begin{figure}[htbp]
    \centering
    \includegraphics[scale=0.5]{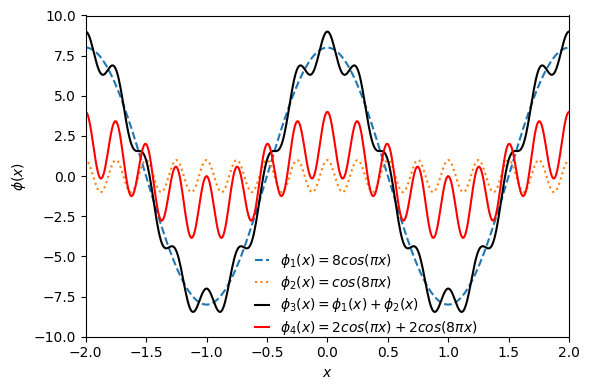}
    \caption{The combination of low frequency and high frequency with different amplitudes. \( \phi_1 (x) \) is a low-frequency function with high amplitude, \( \phi_2 (x) \) is a high-frequency function with low amplitude, \( \phi_3 (x) \) is a combination of \( \phi_1 (x) \) and \( \phi_2 (x) \) to mimic the spectral bias and \( \phi_4 (x) \) is a combination of low-frequency and high-frequency with the same amplitude, an expected output of \( h_2 \).}
    \label{fig:figure2}
\end{figure}

The Fourier bases \( \mathbf{B} \) are weighted by a fixed coefficient matrix \( \mathbf{\Lambda} \in \mathbb{R}^{d_{\text{out}} \times (F \times P)} \). 
The input features are projected onto the weighted Fourier bases to transform the input into a high-dimensional feature vector of global frequency components 
\( \mathbf{h}_1 \in \mathbb{R}^{d_{\text{out}}} \):
\begin{equation}
\mathbf{h}_1 = \mathbf{\Lambda B} \mathbf{x}_{\text{rep}} + \mathbf{b}
\label{eq:13}
\end{equation}
where \( \mathbf{b} \in \mathbb{R}^{d_{\text{out}}} \) is bias, which is the only trainable parameter in Eq. (13). 
The output \( \mathbf{h}_1 \) is a linear combination of the input features \( \mathbf{x}_{\text{rep}} \) weighted by a Fourier-composed filter \( \mathbf{W} = \mathbf{\Lambda B} \). 
By design, these Fourier features span both low and high frequencies, providing a broad spectral representation of the input signal. 
Next, a Gaussian function is applied to the Fourier features to enhance high-frequency resolution as a smooth weighting function,
\begin{equation}
\varepsilon = \exp(-\gamma \mathbf{h}_1^2)
\label{eq:gaussian-factor}
\end{equation}
where \( \gamma \) is a fixed shaping parameter. This Gaussian factor \( \varepsilon \) then multiplies the Fourier features,
\begin{equation}
\mathbf{h}_2 = \varepsilon \mathbf{h}_1
\label{eq:gaussian-output}
\end{equation}
The effect of this multiplication is a soft weighting of \( \mathbf{h_1} \). 
The smooth, non-linear Gaussian factor \( \varepsilon \) attenuates the amplitude of components if the absolute value is large, while leaving small magnitudes nearly unchanged. 
This can be considered as amplitude regularization to prevent any single frequency component from dominating the output. 
The notorious spectral bias that arises in INRs is because the low-frequency features often produce large amplitudes and tend to dominate the early stages of optimization, whereas high-frequency components fluctuate quickly and disappear.
A similar effect has also been observed in quantum models based on data re-uploading circuits. 
The coefficients associated with high-frequency components vanish exponentially with circuit 
depth or number of encoding layers\citep{mhiri2024constrainedvanishingexpressivityquantum}. Consequently, such architectures are 
inherently biased toward smooth, low-frequency functions and struggle to approximate high-frequency structure. 
However, Gaussian scaling can effectively mitigate this issue by penalizing the low-frequency components 
on the corresponding amplitude, giving rise to a balanced frequency distribution. As shown in Figure~\ref{fig:figure2}, 
due to the spectral bias, the fitting function \( \phi_3(x) \) is mainly dominated by low-frequency components, 
leading to a similar distribution as \( \phi_1(x) \). With the Gaussian factor, the predominance of 
low frequency is removed, thus the fitting function \( \phi_4(x) \) includes both low-frequency and 
high-frequency components, providing a more complex structure.

At the end of the FGFS layer, a Linear layer is used to make the model flexible in order to adjust 
the frequency spectrum towards the target domain. A \texttt{BatchNorm} layer is also used to ensure 
training stability and accelerate the convergence of the model. This multi-component FGFS layer provides 
an input embedding that is not only frequency-rich but also spectrally balanced, enabling faster 
convergence and improved representation fidelity.

The output of this classical FGFS layer is used as the input encodings of the quantum circuit.
The quantum circuit is generated by Elivagar\citep{anagolum2024elivagarefficientquantumcircuit}, a noise-aware framework for quantum architecture search. 
The quantum circuit is implemented on 8 qubits, including 16 encoding gates and 256 trainable gates, which contain CZ, \( R_X, R_Y, R_Z \), $\sqrt{X}$ and X quantum gates.
The quantum circuit employs the Super-Parallel ansatz structure\citep{Casas_2023}, where the encoding gates and trainable gate are intertwined inside the quantum circuits to reach the maximum of the degree of freedom (DOF) of the Fourier series. 
After the measurement of the qubits, a linear layer is used to transform the expectation values to the target signal.  

This hybrid model is a nested Fourier composition: the classical layer provides Fourier features, 
and the quantum circuit can further combine them in its own Fourier-expansion output. 
This greatly expands the set of frequencies and frequency combinations that can appear in the final model output, compared to using a raw encoding. 
In addition, this approach avoids having to stack many identical encoding gates in the quantum circuit to reach higher-frequency terms. 
Prior work has noted that to extend a quantum model’s DOF, one can increase the number of qubits or reuploading layers\citep{PhysRevA.103.032430}. 
Here we achieve a similar extension of the frequency spectrum classically: the frequencies are chosen (or learned) to span the range of interest, so the quantum circuit does not need repetitive encodings gates to get those frequencies. 
In other words, the encoding spectrum has been pre-enriched by the classical layer. By diversifying the frequencies at the input, the frequency redundancy of quantum circuits can be effectively mitigated. 
\begin{table}[htbp]
  \centering
  \caption{The performance of different models on image representation. Params is the number of trainable parameters in each model. }
  \label{tab:representation}                            
  \begin{tabular}{llcccccc}
    \toprule
    \multicolumn{2}{c}{\textbf{Model}} & \multicolumn{2}{c}{\textbf{Pneumonia}} & 
    \multicolumn{2}{c}{\textbf{Path}} & 
    \multicolumn{2}{c}{\textbf{Breast}} \\
    \cmidrule(r){1-2} \cmidrule(lr){3-4} \cmidrule(lr){5-6} \cmidrule(l){7-8}
    \textbf{Name} & \textbf{Params} & PSNR $\uparrow$ & SSIM $\uparrow$ & PSNR $\uparrow$ & SSIM $\uparrow$ & PSNR $\uparrow$ & SSIM $\uparrow$ \\
    \midrule
    ReLU        & 841 & 29.495 & 0.965 & 27.978 & 0.955 & 30.158 & 0.932 \\
    Tanh        & 841 & 27.106 & 0.934 & 24.974 & 0.912 & 27.554 & 0.874 \\
    ReLU+RFF    & 791 & 30.533 & 0.973 & 28.227 & 0.962 & 30.654 & 0.940 \\
    SIREN       & 701 & 30.938 & 0.973 & 29.321 & 0.969 & 31.588 & 0.956 \\
    QIREN       & 657 & 31.256 & 0.975 & 29.813 & 0.971 & 32.649 & 0.966 \\
    QFGN        & 585 & 31.927 & 0.980 & 31.173 & 0.978 & 33.372 & 0.970 \\
    \bottomrule
  \end{tabular}
\end{table}
\section{Experiments}
\label{sec: Experiments}
We evaluate the proposed QFGN method in the field of medical imaging, mainly for image representation and image super-resolution. 
We compare it with five baseline models including both quantum and classical neural networks. 
For a fair comparison, we limit the trainable parameters in all models to the same range, around 0.5k\textasciitilde{}0.9k parameters. 
Each model is trained five times, and we report the best performance for comparison purposes. 
For quantum-enabled models, we use TorchQuantum for efficient training on a noiseless simulator while classical neural networks use PyTorch and all models use the Adam optimizer. 
We also do inference on quantum hardware (IBM devices) to achieve real quantum computing. 
In the current NISQ era where noise is unavoidable, we investigate the effect of various error mitigation techniques on QFGN. 
In addition, we compare different quantum hardware. The comprehensive experiments of quantum computing on real hardware provides a clear understanding of QML implemented on contemporary quantum hardware. 
\subsection{Image Representation}
\begin{figure}[htbp]
    \centering
    \includegraphics[width=1\textwidth]{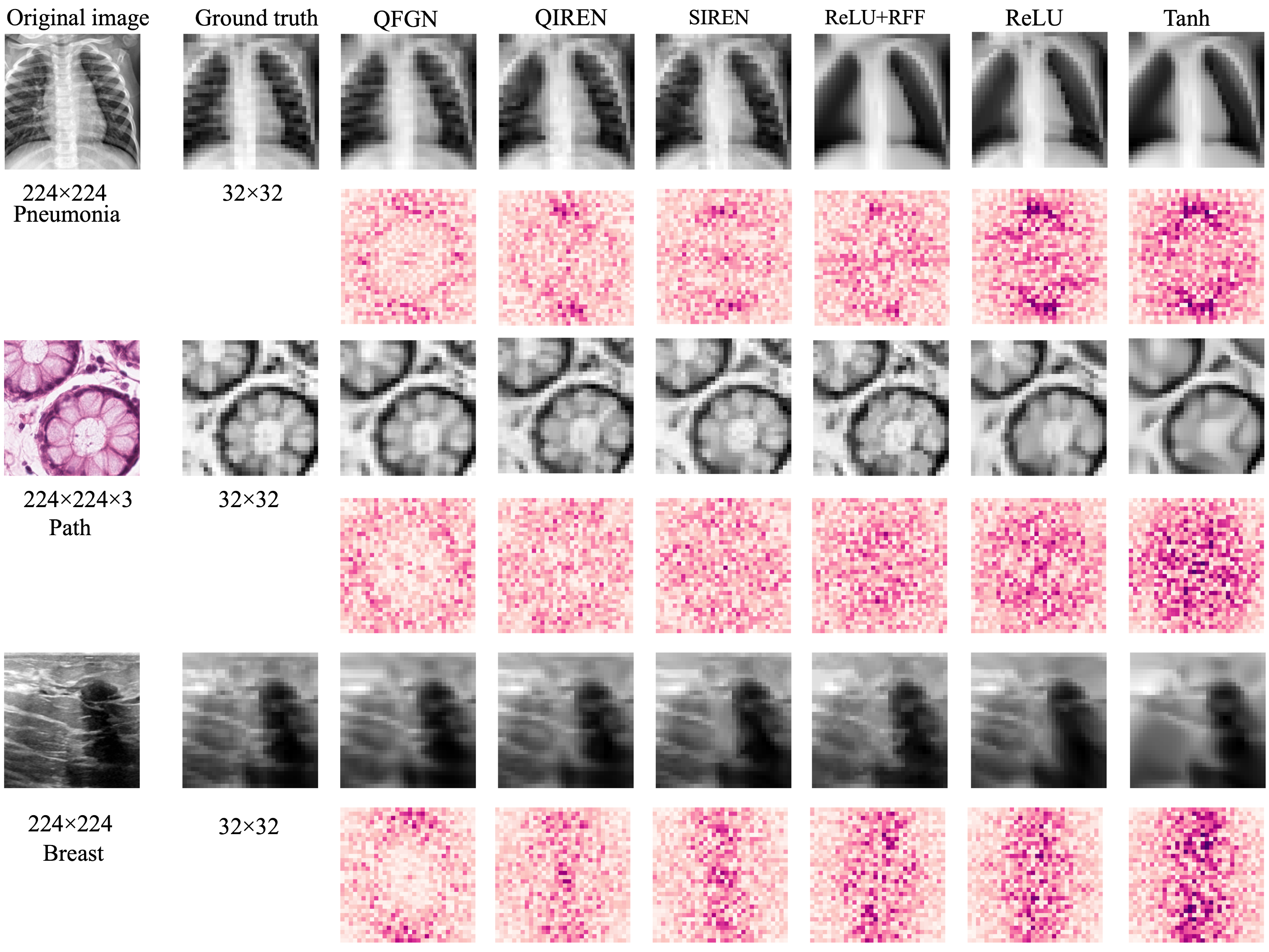}
    \caption{Image representation of each model and the corresponding frequency error with regard to the ground truth. The dark color indicates a large error. }
    \label{fig:figure3}
\end{figure}

In this task, a medical image is considered as an implicit function, mapping its 2D coordinates (x,y) to the corresponding color intensity. 
We consider ReLU-MLP, Tanh-MLP, ReLU-MPL with RFF\citep{tancik2020fourierfeaturesletnetworks}, SIREN\citep{sitzmann2020implicitneuralrepresentationsperiodic} and QIREN\citep{zhao2024quantumimplicitneuralrepresentations} as baseline models. 
Each model is evaluated based on its ability to reconstruct the image. 
We consider three different images from MedMNIST: Pneumonia, Path, Breast.\citep{Yang_2023} 
Pneumonia and Breast images are directly downsampled from 224×224 pixels to 32×32 pixels while the Path image is converted to grayscale and downsampled to a resolution of 32×32. 
The experimental results of all models are reported in Table~\ref{tab:representation}. It is evident that our model, QFGN with the minimal number of parameters, outperforms all other models for both peak signal-to-noise-ratio(PSNR) and structural similarity index measure (SSIM) metrics. 
Notably, with a reduction of 16.5\% of the trainable parameters, QFGN achieves a maximum improvement of 5.6\% in PSNR and a 1.5\% improvement in SSIM when compared to the SOTA SIREN model, 
indicating the intrinsic structure advantage of quantum circuits as Fourier series. 
The image examples and the frequency error are shown in Figure~\ref{fig:figure3}. 
In all three image representation tasks, QFGN consistently achieved the lowest error in capturing various frequency components.

\subsection{Image Super-Resolution}
\begin{figure}[htbp]
    \centering
    \includegraphics[width=1\textwidth]{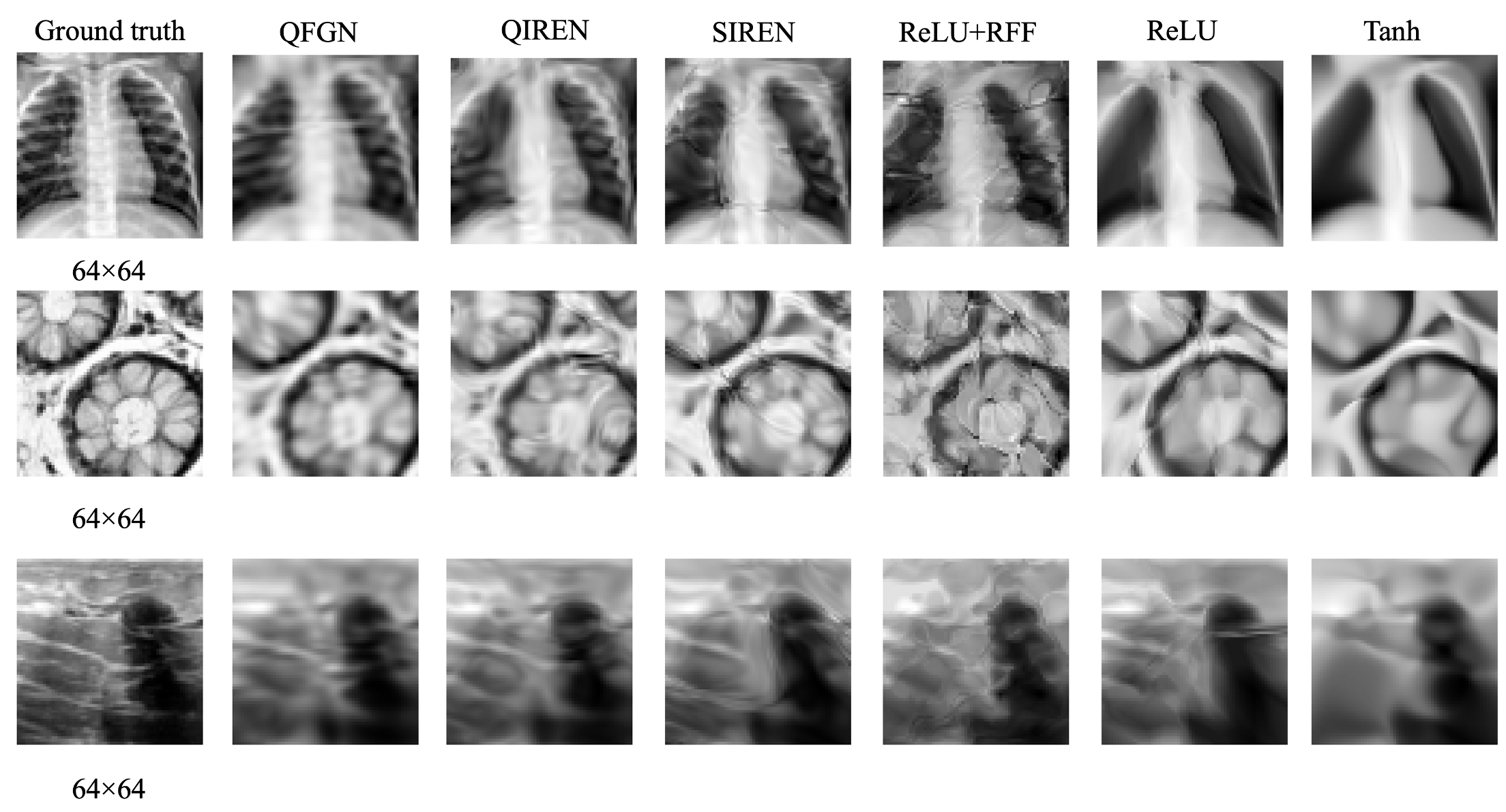}
    \caption{Image super- resolution of each model. }
    \label{fig:figure4}
\end{figure}

To evaluate each trained model in generating high-resolution images, each original image is down sampled to 64×64 pixels as the ground truth and each model is given a grid of 64×64 as inputs to construct 64×64 images. 
The results are reported in Table~\ref{tab:superresolution}. The image examples are given in Figure~\ref{fig:figure4}. 
Our QFGN model consistently surpasses all baseline models. 
It achieves a maximum of a 17.3\% improvement in PSNR and a 22.4\% improvement in SSIM when compared to the classical neural networks, demonstrating this quantum-based model has a superior capability in interpreting implicit functions.
\begin{table}[htbp]
    \centering
    \caption{The performance of different models on image super-resolution. }
    \label{tab:superresolution}
    \begin{tabular}{llcccccc}
      \toprule
      \multicolumn{2}{c}{\textbf{Model}} & \multicolumn{2}{c}{\textbf{Pneumonia}} & 
      \multicolumn{2}{c}{\textbf{Path}} & 
      \multicolumn{2}{c}{\textbf{Breast}} \\
      \cmidrule(r){1-2} \cmidrule(lr){3-4} \cmidrule(lr){5-6} \cmidrule(l){7-8}
      \textbf{Name} & \textbf{Params} & PSNR $\uparrow$ & SSIM $\uparrow$ & PSNR $\uparrow$ & SSIM $\uparrow$ & PSNR $\uparrow$ & SSIM $\uparrow$ \\
      \midrule
      ReLU        & 841 & 24.786 & 0.777 & 22.490 & 0.767 & 24.958 & 0.713 \\
      Tanh        & 841 & 23.727 & 0.702 & 21.309 & 0.701 & 24.086 & 0.636 \\
      ReLU+RFF    & 791 & 23.110 & 0.738 & 20.059 & 0.683 & 24.840 & 0.698 \\
      SIREN       & 701 & 24.602 & 0.806 & 22.368 & 0.804 & 25.676 & 0.766 \\
      QIREN       & 657 & 25.549 & 0.833 & 22.989 & 0.814 & 26.392 & 0.804 \\
      QFGN        & 585 & 25.746 & 0.835 & 23.533 & 0.836 & 26.479 & 0.806 \\
      \bottomrule
    \end{tabular}
  \end{table}
\subsection{QFGN on Quantum Hardware}
All results presented in previous sections are obtained from a noiseless simulator via classical computing. 
However, quantum circuits are designed for execution on quantum computers. 
In view of the current technical limitations in quantum computing, we only do inference on quantum hardware.

In our initial test runs, we only select 100 datapoints from the Breast dataset in image representation task.  
We test various combinations of available techniques, including dynamical decoupling (DD), twirling, Twirled readout error extinction (TREX), zero-noise extrapolation (ZNE), probabilistic error amplification (PEA), and probabilistic error cancellation (PEC). 
The quantum circuit is run on IBM-Kingston (156 qubit Heron-r2) using the Qiskit platform. 
We use Mean Squared Error (MSE) as an evaluation metric.  Results are reported in Table~\ref{tab:error-mitigation}. We notice that these techniques are not always effective, some of which even worsen the performance. 
For example, with DD, the MSE increases by 40.91\% when compared with the performance without any error calibration. 
One possible reason is that the quantum circuit is too dense. 
DD is useful when qubits experience long idle periods, allowing the inserted pulse sequences to counteract decoherence. 
In densely packed circuits where such idle gaps are minimal or absent, the added pulses do not provide meaningful suppression and instead introduce additional sources of error due to pulse imperfections. 
But the combination of DD+TREX+Twirling+ZNE decreases the error by 23.86\%, indicating its effectiveness in our specific case. 
\begin{table}[htbp]
    \centering
    \caption{Error mitigation and suppression results for 100 datapoints.}
    \label{tab:error-mitigation}
  
    \begin{tabular}{|l|c|c|c|c|c|c|}
      \hline
      \textbf{Setting} & No mitigation & DD & \makecell{DD,\\TREX} & \makecell{DD,\\TREX,\\Twirling} & 
      \makecell{DD, TREX,\\Twirling, ZNE} & \makecell{DD, TREX,\\Twirling, ZNE, PEC} \\
      \hline
      MSE & 0.0088 & 0.0124 & 0.0113 & 0.0170 & 0.0067 & 0.0209 \\
      \hline
    \end{tabular}
\end{table}

\begin{figure}[htbp]
    \centering
    \includegraphics[width=1\textwidth]{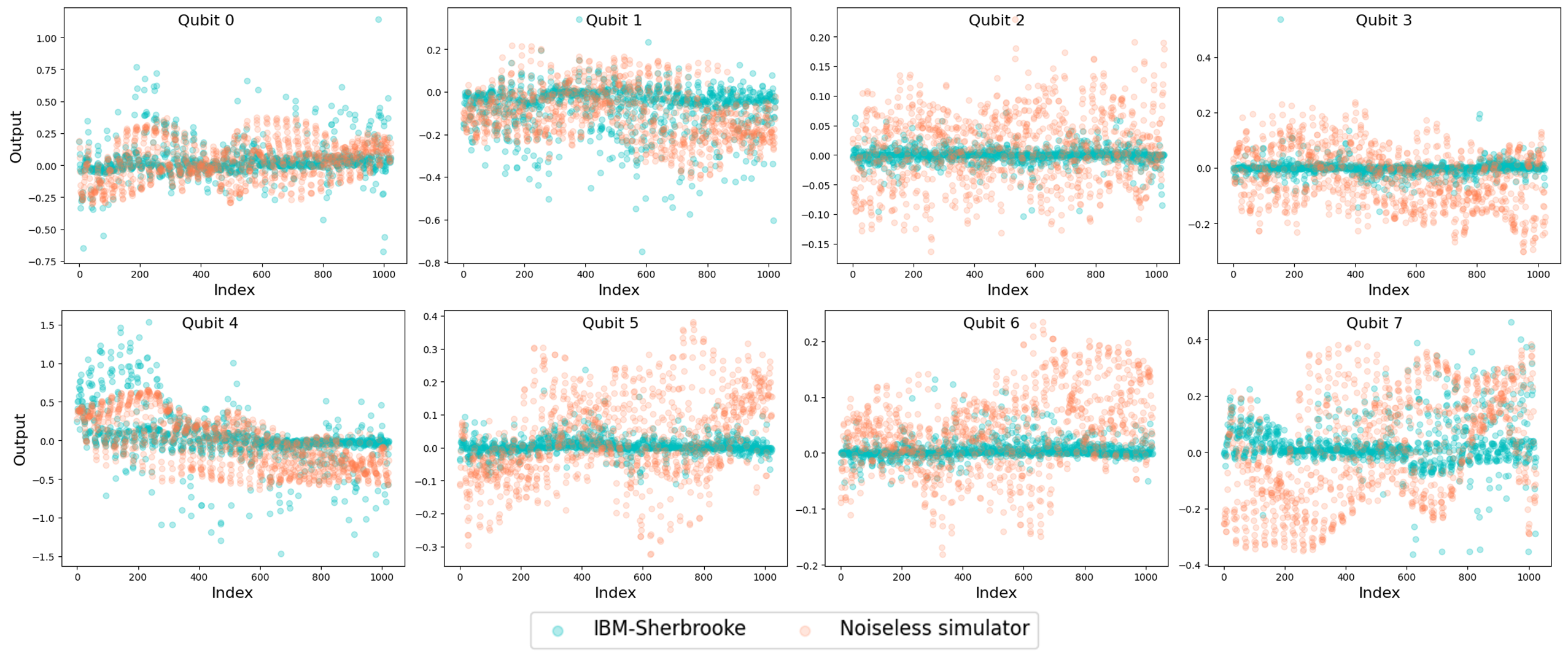}
    \caption{The output from IBM-Sherbrooke. }
    \label{fig:figure5}
\end{figure}

Next, we run the whole dataset (1024 datapoints) on hardware in different settings: No mitigation on IBM-Kingston, DD+TREX+Twirling+ZNE on IBM-Kingston, DD+TREX+Twirling+ZNE on IBM-Sherbrooke (127 qubit Eagle-r3). 
We run the quantum circuit on different IBM devices to investigate how the advancement on hardware can improve the performance. 
The results are presented in Table~\ref{tab:breast_representation}. It is evident that the Heron r2 processor (IBM-Kingston), as expected, outperforms the Eagle r3 processor (IBM-Sherbrooke) in the IBM hardware family. 
We notice that due to the hardware noise, all expectation values obtained from IBM-Sherbrooke fluctuate around zero with little variance(see Figure~\ref{fig:figure5}) , leading to much error.
While with the level of noise on IBM-Kingston, our QFGN model still achieves comparable accuracy to the SIREN model and clearly outperforms all other classical implicit representation networks.  
\begin{table}[htbp]
    \centering
    \caption{The Breast image representation on hardware.}
    \label{tab:breast_representation}
    \begin{tabular}{|l|c|c|c|}
      \hline
      \textbf{Setting} & 
      \makecell{No mitigation\\(IBM-Kingston)} & 
      \makecell{DD, TREX,\\Twirling, ZNE\\(IBM-Kingston)} & 
      \makecell{DD, TREX,\\Twirling, ZNE\\(IBM-Sherbrooke)} \\
      \hline
      PSNR(dB) $\uparrow$ & 23.880 & 31.532 & 14.197 \\
      SSIM $\uparrow$     & 0.808  & 0.955  & 0.270 \\
      \hline
    \end{tabular}
\end{table}

\section{Conclusion}
\label{sec: Conclusion}
In this work, we present QFGN for quantum implicit neural representations. 
The intrinsic structural property of quantum circuits shows the quantum advantages of QFGN over classical neural networks as Fourier series. 
To mitigate the frequency redundancy of quantum circuits, we design a classical module to provide a balanced frequency spectrum derived from Fourier basis functions and Gaussian functions as prior knowledge to enrich the input encodings of the quantum circuits. 
The superior performance of this hybrid QML model on image representation has been validated on real quantum hardware. 
In addition, the model confirms the recent impressive progress in quantum hardware, indicating that noise effects, within the QFGN model, can be substantially reduced on the latest device.  
Overall, this fundamental work provides insights about the theoretical analysis and the broad applications of QML in the near future. 

\section*{Acknowledgements}
The authors gratefully acknowledge financial support from the NIH (GM130641). We also extend our sincere gratitude
to Abdullah Ash Saki (Researcher at IBM Quantum) for his suggestions regarding hardware implementation.
\bibliographystyle{unsrtnat}
\bibliography{neurips_2025}

\newpage
\appendix

\renewcommand{\thefigure}{A.\arabic{figure}}
\setcounter{figure}{0}

\section{Additional details of quantum circuit}
Quantum computing is based on qubits, which can exist in superpositions of states, enabling quantum computers to process a vast number of possibilities simultaneously. In this framework, information is stored in the quantum state of a system, typically described as a vector in a complex Hilbert space. A quantum state, such as \( |\psi\rangle \), encodes the probability amplitudes of all possible classical configurations of the qubits. The evolution and manipulation of this quantum information are governed by quantum gates—unitary operations that preserve the norm of the state vector. Single-qubit gates, such as the Pauli gates \( X, Y, Z \), the Hadamard gate \( H \), and rotation gates \( R_X(\theta), R_Y(\theta), R_Z(\theta) \), are fundamental building blocks that control the behavior of individual qubits by altering their superposition or phase. These gates enable precise manipulation of the quantum state, forming the basis for more complex multi-qubit operations and quantum algorithms.

After the manipulation of qubits through quantum gates, the final step in a quantum circuit involves \textit{measurement}, which collapses the quantum state into a classical outcome. Measurement projects the quantum state \( |\psi\rangle \) onto a basis state—typically the computational basis—yielding a binary result that reflects the probability distribution encoded in the quantum amplitudes. This process irreversibly extracts classical information from the quantum system. In practice, repeated measurements of identically prepared quantum states are used to estimate expectation values of observables, such as \( \langle \psi | O | \psi \rangle \), where \( O \) is a Hermitian operator corresponding to a measurable quantity. These classical outcomes serve as the output of quantum algorithms and are essential for interpreting results from quantum computations.

The quantum circuit of QFGN is composed of CZ, \( R_X, R_Y, R_Z \), $\sqrt{X}$ and X quantum gates, which are defined as 
\[
\text{CZ} =
\begin{pmatrix}
1 & 0 & 0 & 0 \\
0 & 1 & 0 & 0 \\
0 & 0 & 1 & 0 \\
0 & 0 & 0 & -1
\end{pmatrix}, \quad
R_X(\theta) =
\begin{pmatrix}
\cos\frac{\theta}{2} & -i\sin\frac{\theta}{2} \\
-i\sin\frac{\theta}{2} & \cos\frac{\theta}{2}
\end{pmatrix},
\]
\[
R_Y(\theta) =
\begin{pmatrix}
\cos\frac{\theta}{2} & -\sin\frac{\theta}{2} \\
\sin\frac{\theta}{2} & \cos\frac{\theta}{2}
\end{pmatrix}, \quad
R_Z(\theta) =
\begin{pmatrix}
1 & 0 \\
0 & e^{i\theta}
\end{pmatrix}, \quad
X =
\begin{pmatrix}
0 & 1 \\
1 & 0
\end{pmatrix},
\]
\[
\sqrt{X} =
\frac{1}{2}
\begin{pmatrix}
1 + i & 1 - i \\
1 - i & 1 + i
\end{pmatrix}
\]

The whole circuit structure is shown in Figure A.1. It includes two types of gates: the trainable gate, $\mathrm{t}[i]$, $i \in \{0, 1, 2, \cdots, 255\}$ and the encoding gate, $\mathbf{x}[k]$, $k \in \{0, 1, 2, \cdots, 15\}$.

\begin{figure}[ht]
    \centering
    \includegraphics[width=1\textwidth]{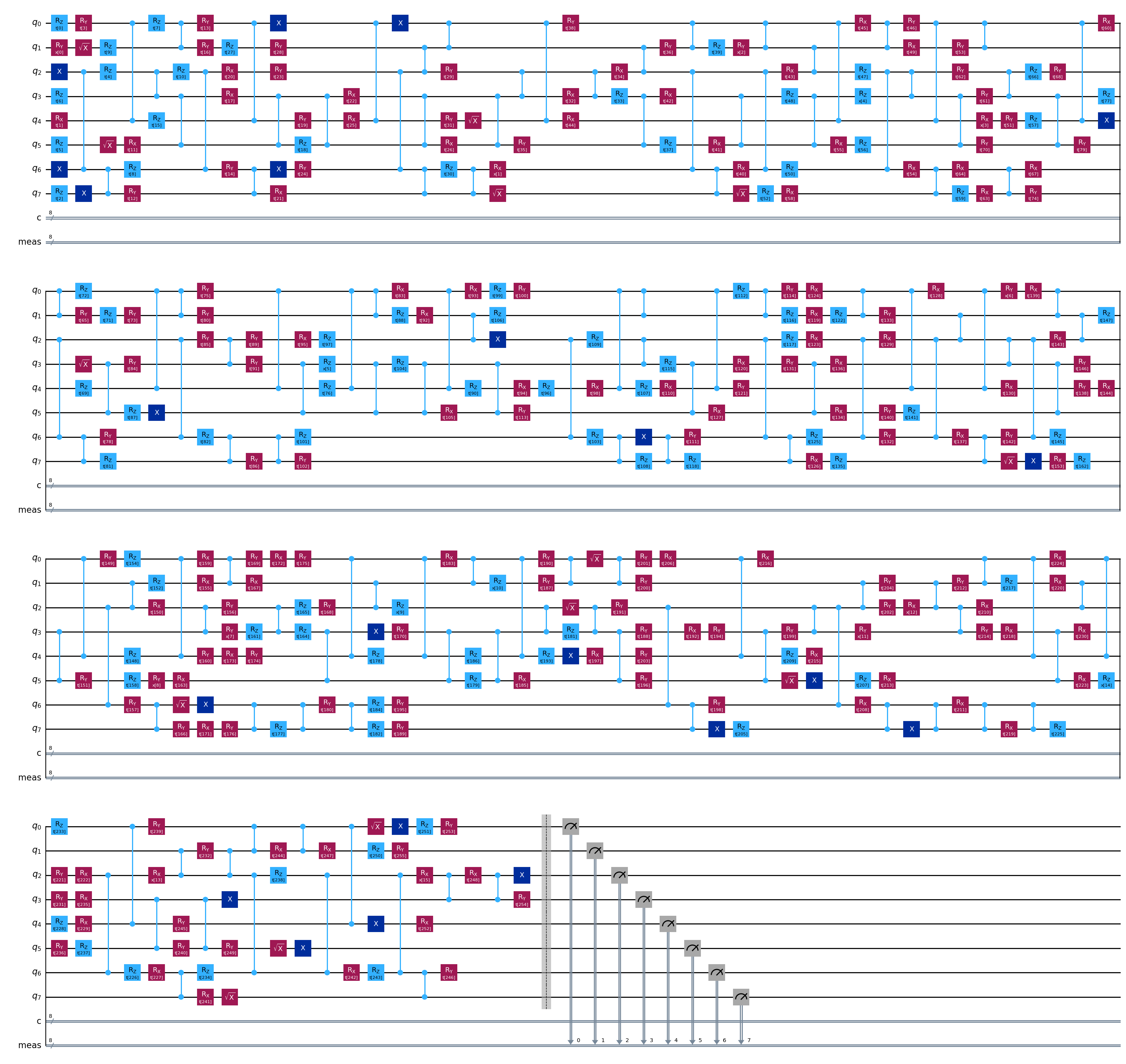}
    \caption{The quantum circuit in QFGN. }
    \label{fig:figureA1}
\end{figure}
\section{Experiment details}
\label{sec: Experiment details}
All models in this work are trained for 600 epochs. The MSE is used as the loss function, and the Adam optimizer is used with the parameters $\beta_1 = 0.9$, $\beta_2 = 0.999$ and $\epsilon = 1\mathrm{e}{-8}$. All models are trained using 8 CPUs via high-performance computing(HPC). The training of classical neural networks usually takes several seconds, while QIREN and QFGN require approximately 5 minutes. 

\subsection{Classical neural networks}
\textbf{ReLU-based MLP} consists of one input linear layer, six hidden linear layers (each with a hidden dimension of 10), and one output linear layer. All layers except the output are followed by a BatchNorm layer and an activation function.

\textbf{Tanh-based MLP} follows the same architecture as the ReLU-based MLP, but uses the Tanh activation function.

\textbf{ReLU-based MLP with Random Fourier Features} shares the same architecture as the ReLU-based MLP, with the exception that the input layer is replaced by a RFF layer. This RFF layer includes a non-trainable random parameter matrix followed by sine and cosine activations.

\textbf{SIREN} mirrors the ReLU-based MLP architecture but replaces BatchNorm layers with a specialized initialization scheme and uses sine as activation function.
\subsection{Quantum models}
\textbf{QIREN} comprises three hybrid layers and one output linear layer. Each hybrid layer integrates a linear layer (hidden dimension of 8), a BatchNorm layer, and a quantum circuit operating on 8 qubits.

\textbf{QFGN} comprises three modules. The first module integrates a Fourier function, a Gaussian function($\gamma=0.8$), a linear layer with a hidden dimension of 16, and a BatchNorm layer. The second module consists of a quantum circuit with 8 qubits, followed by a linear layer with dimensions $(8, 1)$.

\subsection{Hardware implementation}
We use Qiskit to parameterize the quantum circuit by assigning each trained parameter to its corresponding quantum gate. The Estimator primitive is employed to obtain the expectation value of each qubit, using 50,000 shots.

\section{Representation capacity: quantum circuit vs.  ReLU-Based MLPs with Fourier features}
\label{sec: cnn}
Standard ReLU-activated MLPs are known to struggle with representing high-frequency content – their spectral amplitude decays rapidly at higher frequencies. To address this, recent Fourier Neural Networks (FNNs) augment or replace the ReLU network with sinusoidal representations, either by encoding inputs with trigonometric functions or using sinusoidal activations. In essence, these approaches allow the network to learn a truncated Fourier series expansion of the target function. We outline the mathematical derivation below.

\textbf{Random Fourier Feature (RFF) Mapping:} Consider a single-layer perceptron that first maps its input 
$\mathbf{x} \in \mathbb{R}^{d_{\text{in}}}$ through a fixed Fourier feature embedding 
$\gamma(\mathbf{x})$ and then applies a linear layer. This model can be written as:

\[
\mathbf{y}(\mathbf{x}) = \mathbf{W} \, \gamma(\mathbf{x}) + \mathbf{b},
\]

where $\mathbf{W} \in \mathbb{R}^{d_{\text{out}} \times 2m}$ and 
$\mathbf{b} \in \mathbb{R}^{d_{\text{out}}}$ are the output weights and biases. 
The feature map $\gamma(\mathbf{x})$ stacks sinusoidal functions:

\[
\gamma(\mathbf{x}) = 
\begin{pmatrix}
\cos(2\pi \mathbf{B} \mathbf{x}) \\
\sin(2\pi \mathbf{B} \mathbf{x})
\end{pmatrix},
\]

with $\mathbf{B} \in \mathbb{R}^{m \times d_{\text{in}}}$ being a matrix of $m$ frequency vectors 
(often chosen at random). Here $m$ — the number of frequency components — determines the size of this 
mapping and thus serves as a measure of the model’s capacity.

Intuitively, $\gamma(\mathbf{x})$ produces $m$ pairs of cosine and sine features, 
$\cos(2\pi \mathbf{b}_k^\top \mathbf{x})$ and $\sin(2\pi \mathbf{b}_k^\top \mathbf{x})$ 
for $k = 1, \dots, m$, using the $k$-th row $\mathbf{b}_k$ of $\mathbf{B}$ as the frequency. 
The perceptron’s output is then a linear combination of these sinusoidal basis functions.

\textbf{Fourier Series Background:} If the input domain for $\mathbf{x}$ is bounded (e.g. $\mathbf{x} \in [0,1]^d$), one can assume without loss of generality that the target signal $f(\mathbf{x})$ is extended periodically outside this domain. Under this assumption, $f$ admits a Fourier series expansion. In $d$ dimensions (assuming period 1 in each dimension), this expansion is:

\[
f(\mathbf{x}) = \sum_{\mathbf{n} \in \mathbb{Z}^d} c_{\mathbf{n}} \, e^{2\pi i\, \mathbf{n} \cdot \mathbf{x}},
\]

with complex coefficients $c_{\mathbf{n}}$ given by the usual Fourier integral formula. For real-valued signals, the coefficients obey $c_{-\mathbf{n}} = c_{\mathbf{n}}^*$ (complex conjugate symmetry), which means the series can be expressed in terms of real sine and cosine terms. In the one-dimensional case, for example, the real form of the Fourier series (truncated to frequencies $|k|\le N$) is:

\[
f(x) \approx a_0 + \sum_{k=1}^{N} \Big[ a_k \cos(2\pi k x) + b_k \sin(2\pi k x) \Big]. \tag{*}
\]

Each pair $(a_k, b_k)$ corresponds to the Fourier coefficients at frequency $k$ (related to $c_k$ and $c_{-k}$). As the number of terms $N$ grows, this series can approximate any band-limited or continuous periodic function arbitrarily well.

\textbf{Equivalence to a Truncated Series:} Now, comparing the neural network’s output to the Fourier series form reveals a clear correspondence. The RFF network output can be expanded as:

\[
y_j(\mathbf{x}) = \sum_{k=1}^{m} W^{(c)}_{j,k} \cos(2\pi \, \mathbf{b}_k \cdot \mathbf{x}) + \sum_{k=1}^{m} W^{(s)}_{j,k} \sin(2\pi \, \mathbf{b}_k \cdot \mathbf{x}) + b_j,
\]

for each output dimension $j = 1, \dots, d_{\text{out}}$, where $W^{(c)}$ and $W^{(s)}$ are the portions of $\mathbf{W}$ applied to the cosine and sine features of $\gamma(\mathbf{x})$.

This is exactly a \textbf{truncated Fourier series} representation of $f_j(x)$, containing $m$ sinusoidal terms (plus a bias). In other words, the network’s learnable parameters $\mathbf{W}$ supply the Fourier series coefficients, and the chosen rows of $\mathbf{B}$ determine which frequencies appear in the series. 

Thus, a ReLU MLP augmented with an $m$-dimensional Fourier feature embedding is effectively \textbf{learning the coefficients of a Fourier series truncated to $m$ frequency components}.

\subsection*{Sine-Activated Networks as Learned Fourier Series}

An alternative way to incorporate Fourier structure is to use sinusoidal activations in the network itself. \textbf{Sinusoidal Representation Networks (SIREN)} replace ReLU with sine activations, allowing each neuron to produce a sinusoidal output. For example, a one-hidden-layer SIREN with weights $\mathbf{W}_1$, $\mathbf{W}_2$ and biases $\mathbf{b}_1$, $\mathbf{b}_2$ can be written as:

\[
\mathbf{y}(\mathbf{x}) = \mathbf{W}_2 \, \sin(\mathbf{W}_1 \mathbf{x} + \mathbf{b}_1) + \mathbf{b}_2.
\]

This architecture can also be viewed through a Fourier lens. In fact, one can rewrite the earlier RFF perceptron in a form with a single sinusoidal activation by absorbing cosine terms into a phase shift. Using the identity $\cos\theta = \sin(\theta + \pi/2)$, we can combine the cosine and sine features of $\gamma(\mathbf{x})$ into a single sine term with an offset. This yields an equivalent form:

\[
\mathbf{y}(\mathbf{x}) = \mathbf{W} \, \sin(2\pi \, \mathbf{C} \mathbf{x} + \boldsymbol{\phi}) + \mathbf{b},
\]

for appropriate $\mathbf{C} \in \mathbb{R}^{m \times d_{\text{in}}}$ and phase vector $\boldsymbol{\phi} \in \mathbb{R}^{2m}$. Here, $\mathbf{C}$ plays the role of the first-layer weight matrix and $\boldsymbol{\phi}$ acts as an initial phase bias applied inside the sine.

This is exactly the form of a one-layer SIREN; indeed, it shows that a \textbf{Fourier-mapped perceptron is structurally equivalent to a SIREN with one hidden layer}. The only difference is that in SIREN, the frequency matrix $\mathbf{C}$ (analogous to $\mathbf{B}$) and phase $\boldsymbol{\phi}$ are \textit{trainable parameters} rather than fixed features.

In other words, SIREN learns which Fourier basis frequencies to use, instead of relying on a predetermined (e.g. random or fixed lattice) encoding. Both the RFF approach and SIREN ultimately produce networks whose \textbf{fundamental building blocks are sinusoidal functions}, which is why both are referred to as Fourier Neural Networks.

Viewing ReLU-based MLPs with Fourier features or sine activations as learning a truncated Fourier series provides a clear theoretical insight: the representational capacity grows \textbf{linearly} with the number of sinusoidal terms (features or neurons). However, as indicated in Section~\ref{sec:qc as Fourier series}, the number of frequency components represented by quantum circuit can grow \textbf{exponentially} under optimal conditions where the difference of eigenvalue sums is unique. Therefore, quantum implicit neural representations demonstrate a clear quantum advantage over classical implicit neural representations.

\newpage


\end{document}